\documentclass[prd,amsmath,superscriptaddress,twocolumn,nofootinbib]{revtex4-2}
\pdfoutput=1
\usepackage[utf8]{inputenc}
\usepackage{graphicx}
\usepackage{float}
\usepackage{epsf,latexsym,bbm,euscript}
\usepackage{amssymb,amsmath}
\usepackage{mathtools} 
\usepackage{times}
\usepackage{soul,xcolor}
\usepackage{mathtools}
\usepackage{mathrsfs}
\usepackage{array}
\usepackage{booktabs}
\usepackage{enumitem}

\usepackage[caption=false]{subfig}
\newcommand{\subfigimg}[3][,]{%
	\setbox1=\hbox{\includegraphics[#1]{#3}}%
	\leavevmode\rlap{\usebox1}%
	\rlap{\hspace*{-18pt}\raisebox{.5\baselineskip}{\small{#2}}}%
	\phantom{\usebox1}%
}

\def\sg{\textsl{g}}
\newcommand{\RN}[1]{\uppercase\expandafter{\romannumeral #1\relax}}

\usepackage{environ}
\NewEnviron{thincases}{\scalebox{0.5}[1]{$\displaystyle
		\left\{\scalebox{2}[1]{\setlength{\arraycolsep}{6pt}
			$\displaystyle\begin{array}{ll}
				\BODY
			\end{array}$}\right.$}}
			\NewEnviron{thinpmatrix}{\scalebox{0.5}[1]{$\displaystyle
	\left(\scalebox{2}[1]{$\displaystyle
		\begin{matrix}
			\BODY
		\end{matrix}$}
	\right)$}}

\makeatletter
\g@addto@macro\bfseries{\boldmath}
\makeatother

\makeatletter
\newcommand*{\defeq}{\mathrel{\rlap{%
			\raisebox{0.3ex}{$\m@th\cdot$}}%
		\raisebox{-0.3ex}{$\m@th\cdot$}}%
	=}
\newcommand*{\eqdef}{=\mathrel{\rlap{%
			\raisebox{0.3ex}{$\m@th\cdot$}}%
		\raisebox{-0.3ex}{$\m@th\cdot$}}%
}
\makeatother

\def\sg{\textsl{g}}

\usepackage{pifont}
\newcommand{\cmark}{\text{\ding{51}}}
\newcommand{\xmark}{\text{\ding{55}}}

\usepackage{scalerel}
\usepackage{tikz}
\usetikzlibrary{svg.path}
\definecolor{orcidlogocol}{HTML}{A6CE39}
\tikzset{
	orcidlogo/.pic={
		\fill[orcidlogocol] svg{M256,128c0,70.7-57.3,128-128,128C57.3,256,0,198.7,0,128C0,57.3,57.3,0,128,0C198.7,0,256,57.3,256,128z};
		\fill[white] svg{M86.3,186.2H70.9V79.1h15.4v48.4V186.2z}
		svg{M108.9,79.1h41.6c39.6,0,57,28.3,57,53.6c0,27.5-21.5,53.6-56.8,53.6h-41.8V79.1z M124.3,172.4h24.5c34.9,0,42.9-26.5,42.9-39.7c0-21.5-13.7-39.7-43.7-39.7h-23.7V172.4z}
		svg{M88.7,56.8c0,5.5-4.5,10.1-10.1,10.1c-5.6,0-10.1-4.6-10.1-10.1c0-5.6,4.5-10.1,10.1-10.1C84.2,46.7,88.7,51.3,88.7,56.8z};
	}
}
\newcommand\orcidlink[1]{\href{https://orcid.org/#1}{\mbox{\scalerel*{
				\begin{tikzpicture}[yscale=-1,transform shape]
					\pic{orcidlogo};
				\end{tikzpicture}
			}{X}}}}

\usepackage{url,hyperref}
\hypersetup{colorlinks,linkcolor={blue!55!black},citecolor={red!50!black},urlcolor={blue!45!black},breaklinks=true}

\begin{document}

\title{Kinematic and energy properties of dynamical regular black holes}
	
\author{Sebastian Murk\orcidlink{0000-0001-7296-0420}}
\email{sebastian.murk@oist.jp}
\affiliation{Okinawa Institute of Science and Technology, 1919-1 Tancha, Onna-son, Okinawa 904-0495, Japan}

\author{Ioannis Soranidis\orcidlink{0000-0002-8652-9874}}
\email{ioannis.soranidis@hdr.mq.edu.au}
\affiliation{School of Mathematical and Physical Sciences, Macquarie University, Sydney, New South Wales 2109, Australia}

\begin{abstract}
Nonsingular black holes have received much attention in recent years as they provide an opportunity to avoid the singularities inherent to the mathematical black holes predicted by general relativity. Based on the assumption that semiclassical physics remains valid in the vicinity of their horizons, we derive kinematic properties of dynamically evolving spherically symmetric regular black holes. We review the Hawking--Ellis classification of their associated energy-momentum tensors and examine the status of the null energy condition in the vicinity of their horizons as well as their interior. In addition, we analyze the trajectory of a moving observer, find that the horizons can be crossed on an ingoing geodesic, and thus entering and exiting the supposedly trapped spacetime region is possible. We outline the ramifications of this result for the information loss problem and black hole thermodynamics. Throughout the article, we illustrate relevant features based on the dynamical generalization of the regular black hole model proposed in {\href{https://doi.org/10.1007/JHEP09(2022)118}{J.\ High Energy Phys.\ \textbf{09}, 118 (2022)}} and elucidate connections to the only self-consistent dynamical physical black hole solutions in spherical symmetry.
\end{abstract}

\maketitle
	
\section{Introduction} \label{sec:intro}
While the existence of dark massive compact objects has been established beyond any reasonable doubt, their precise physical nature remains enigmatic, with a range of possibilities under consideration \cite{v:08,h:14,f:14,bmt:17,cp:19,m:23}. The prevalent astrophysical description is the Schwarzschild/Kerr black hole paradigm, which is based on the mathematical black hole (MBH) solutions of general relativity (GR). Their hallmark features are the presence of an event horizon and central singularity. Despite the successes of the paradigm, both features that distinguish MBHs are accompanied by empirical and conceptual pathologies which are absent by design in many alternative models describing dark ultracompact objects (UCOs) that are also compatible with observational data.

By its very definition, the event horizon is a global geometric property and thus physically unobservable \cite{v:14}. While current data is consistent with having the Schwarzschild/Kerr solutions as asymptotic final states of gravitational collapse, such objects are \textit{de facto} horizonless for distant observers (as their horizons exist only for $t\to\infty$). In contrast, physical black holes (PBHs) \cite{mmt:rev:22} bounded by a dynamically evolving quasilocal horizon formed in finite time according to the clock of a distant observer are (at least in principle) physically observable, i.e.\ there is a measurement that can be performed in a finite time interval and within a finite-size region of spacetime to determine the presence or absence of a quasilocal horizon. The presence of a physical singularity (i.e.\ one that is not an artifact of a particular choice of coordinates) inevitably introduces nontrivial causal structures into the spacetime at large \cite{p:65} and is typically interpreted as a harbinger that the underlying theory breaks down\footnote{``The crushing of matter to infinite density by infinite tidal gravitational forces is a phenomenon with which one cannot live comfortably. [...] it is difficult to believe that physical singularities are a fundamental and unavoidable feature of our universe. [...] one is inclined to discard or modify that theory rather than accept the suggestion that the singularity actually occurs in nature.'' \cite{t:66}; ``[...] when Relativity and Quantum Mechanics are melded it will be shown that there are no singularities anywhere. When theory predicts singularities, the theory is wrong!'' \cite{k:23}}. An immediate consequence that is frequently discussed in the literature is the so-called information loss paradox \cite{h:16}: in this scenario, the apparent lack of unitarity in the black hole evaporation process is typically ascribed to the propagation of quantum correlations into the singularity located at the black hole's interior \cite{w:01}. To maintain unitarity and avoid information loss, it has been conjectured that the time dependence of the Hawking radiation's entanglement entropy follows the Page curve \cite{p:93,p:13}, i.e.\ it first increases until it reaches its peak at the Page time (where it coincides with the Bekenstein--Hawking entropy \cite{b:72,b:73,b:74,h:74,h:75,h:76} of the black hole) and subsequently decreases until it reaches zero at the end of the evaporation process (corresponding to a final pure state)\footnote{The results of a recent study \cite{cos:23} indicate that the evaporation of two-dimensional nonsingular dilatonic black holes, which can be used to model some of the thermodynamic properties of four-dimensional nonsingular black holes, conforms to the unitary evolution predicted by the Page curve.}.
	
Several alternatives to MBHs have been proposed to describe the observed astrophysical black hole candidates and avoid the presence of singularities, including horizonless UCOs \cite{cp:19}, gravastars \cite{mm:04,mm:23}, wormholes \cite{e:73,mt:88,sv:19}, and fuzzballs \cite{lm:02,m:05}. Our analysis in this article focuses on dynamical models of ``horizonful'' but singularity-free regular black holes (RBHs) \cite{b:68,d:92,h:06,b:book:23} that are bounded by a quasilocal inner and outer horizon. In such models, the spacetime is typically regularized through the introduction of a minimal length scale that arises from a theory of quantum gravity and acts as a Planckian cutoff. In spherical symmetry, the central singularity is replaced by a de Sitter core and the introduction of a minimal length scale leads to the emergence of an inner horizon \cite{f:16}. The minimal length scale is bounded from above by observational data, such as the trajectory of the S2 star orbiting Sagittarius A$^\star$ \cite{cdddos:23}, i.e.\ the astrophysical black hole candidate at the center of the Milky Way galaxy. The close contact with astrophysical observations of dark UCOs highlights the importance of studying realistic candidate models such as dynamically evolving RBHs more thoroughly.

It is worth noting that significant challenges arise from the presence of an inner horizon due to the fact that they are typically unstable under small perturbations. These instabilities are characterized by an exponential growth of the gravitational energy in the neighborhood of the inner horizon (which can be checked by tracking the relevant curvature scalars), a problem known as mass inflation instability \cite{pi:89,pi:90,o:91,ha:10,cdlpv:21,dmt:22}. We use the dynamical generalization of the so-called inner-extremal RBH model proposed in Ref.~\citenum{cdlpv:22} to illustrate the key aspects of our analysis. In this model, the mass inflation problem is resolved at the expense of a degenerate inner horizon with vanishing surface gravity, although a recent study argues that even in this case the mass inflation instability cannot be avoided once semiclassical effects are taken into account \cite{McM:23}. Nevertheless, we stress that the properties we derive in this article are generic and apply to any and all dynamical RBH models described by a metric function of the form given in Eq.~\eqref{eq:genRBH.fvr}.

The absence of the central singularity is the principal characteristic of nonsingular black holes and offers a potential resolution to the information loss problem. In such models, the infalling matter never disappears from the manifold, and may possibly escape. By considering dynamical RBH models rather than their static counterparts, we explore this phenomenon based on the assumption that semiclassical gravity remains valid in their vicinity of their horizons.

The remainder of this article is organized as follows: in Sec.~\ref{sec:RBHs}, we review the construction of spherically symmetric dynamical RBHs in semiclassical gravity from a model-agnostic point of view, and introduce the inner-extremal RBH model we use to illustrate relevant physical features in Figs.~\ref{fig:NEC}--\ref{fig:w1}. We then focus on the classification of energy-momentum tensors (EMTs) describing dynamical RBH metrics, discuss implications for the associated matter content, and in particular for the status of the null energy condition (NEC) in the vicinity and interior of RBHs (Sec.~\ref{sec:EMT.NEC}). In Sec.~\ref{sec:particle.motion}, we consider the perspective of a moving observer attempting to escape the trapped spacetime region. Based on an expression for the linear coefficient of the Misner--Sharp (MS) mass, we highlight the relevance of our results for black hole thermodynamics and the transition between self-consistent dynamical PBH solutions (Sec.~\ref{sec:transition}). Lastly, we summarize the physical implications of our results and comment on possible avenues for future research related to nonsingular black hole spacetimes (Sec.~\ref{sec:concl}). Throughout this article, we use the metric signature $(-,+,+,+)$ and work in dimensionless units such that $c=G=\hbar=k_{B}=1$.

\section{Dynamical regular black holes in semiclassical gravity} \label{sec:RBHs}
\begin{figure}[htbp]
	\includegraphics[scale=0.725]{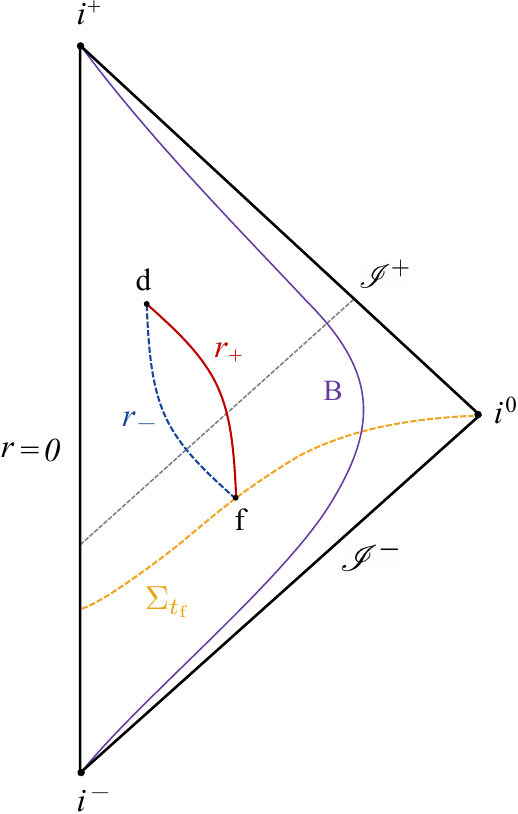}
	\caption{Schematic Carter--Penrose diagram depicting the formation and evaporation of a RBH with an apparent horizon $r_\sg$ that has a shrinking inner $r_-$ (dashed blue line) and shrinking outer $r_+$ (solid red line) component. The trajectory of a distant observer Bob is indicated in purple and marked by the initial $\mathrm{B}$. The dashed gray line corresponds to an outgoing radial null geodesic that reaches future null infinity $\mathscr{I}^{+}$. The asymptotic structure of a simple RBH spacetime coincides with that of Minkowski spacetime. An immediate neighborhood of $r=0$ never belongs to the trapped region. The points $\mathrm{f}$ and $\mathrm{d}$ represent the events of formation and disappearance of the trapped region. The equal time hypersurface $\Sigma_{t_{\mathrm{f}}}$ is shown as a dashed orange line connecting the center $r=0$ and spacelike infinity $i^0$. Both the outer and inner horizon are timelike membranes over the course of the lifespan of the RBH, i.e.\ $\forall v \in (v_\mathrm{f},v_{\mathrm{d}})$.} 
	\label{fig:RBH}
\end{figure}

The geometry of a general spherically symmetric spacetime is described by the line element 
\begin{align}
	ds^2 = - e^{2h_{+}(v,r)} f(v,r) dv^2 + 2 e^{h_{+}(v,r)} dv dr + r^2 d\Omega^2 ,
	\label{eq:gen.metr}
\end{align}
where $d\Omega^2$ denotes the normalized spherically symmetric Riemannian metric on the 2-sphere $S^2$. The metric function $f(v,r)$ is related to the MS mass \cite{ms:64} $M(v,r) = C(v,r)/2$ via
\begin{align}
	f(v,r) \defeq \partial_\mu r \partial^\mu r =1 - \frac{C(v,r)}{r} ,
	\label{eq:fvr.MSmass}
\end{align}
where $r$ denotes the areal radius, the MS mass\footnote{While the MS mass is technically $C(v,r)/2$, we take the liberty to refer to $C(v,r)$ itself as the MS mass as well when there is no need (in terms of its physical relevance) to account for the factor of one half.} is described by the series expansion
\begin{align}
	C(v,r) &= r_+(v) + \sum\limits_{i=1}^\infty w_i(v) \big(r-r_+(v)\big)^i , \label{eq:MSmass}
\end{align}
with coefficients
\begin{align}
	w_i(v) &= \frac{1}{i!} \frac{\partial^i C(v,r)}{\partial r^i} \Big\vert_{r_+(v)} ,
\end{align}
and the functions $h(t,r)$ and $h_+(v,r)$ play the role of integrating factors that turn the expression
\begin{align}
	dt &= e^{-h(t,r)} \left(e^{h_+(v,r)}dv - \frac{dr}{f(v,r)} \right)
\end{align}
into an exact differential (provided that the coordinate transformation exists). Since $h_+(v,r)=0$ in all spherically symmetric RBH spacetimes that we are aware of, we limit our considerations to this case in what follows\footnote{Note also that $h_+(v,r)=0$ at the outer horizon $r=r_+(v)$. For $h_+(v,r)=0$ and $C(v,r) \equiv C(v)$, the metric of Eq.~\eqref{eq:gen.metr} reduces to the Vaidya metric.}. The general spherically symmetric metric specified in Eq.~\eqref{eq:gen.metr} then simplifies to
\begin{align}
	ds^2 = - f(v,r) dv^2 + 2dv dr + r^2 d\Omega^2 ,
	\label{eq:simpl.metr}
\end{align}
and a dynamically evolving RBH is described by a metric function of the form
\begin{align}
	f(v,r) = g(v,r)\big(r-r_{-}(v)\big)^a\big(r-r_{+}(v)\big)^b ,
	\label{eq:genRBH.fvr}
\end{align}
where $r_{-}(v)$ and $r_{+}(v)$ denote the inner and outer horizon, respectively, and $a,b \in \mathbb{N}_{>0}=\lbrace1,2,...\rbrace$ are positive integers labeling their degeneracy. The role of the \textit{a priori} undetermined function $g(v,r)$ is to ensure regularity at the center as well as the proper asymptotic behavior of the metric at infinity \cite{ms:23}. It must be positive for the entire existence of the trapped region, that is $g(v,r)>0$ for every $v_\mathrm{f} \leqslant v \leqslant v_\mathrm{d}$, where $v_\mathrm{f}$ and $v_\mathrm{d}$ denote the formation and disappearance of the trapped region, respectively, as indicated in Fig.~\ref{fig:RBH}.

The expansions of ingoing and outgoing radial null geodesics are given by
\begin{align}
	\theta_{-}=-\frac{2}{r} \; , \quad \theta_{+}=\frac{f(v,r)}{r} \; ,
	\label{eq:rad.null.geod.exp}
\end{align}
respectively. The existence of a trapped spacetime region is determined by the signature of their product $\theta_{-} \theta_{+} \stackrel{?}{\lessgtr} 0$. Following the argumentation in Ref.~\citenum{h:94}, a trapped region is present if and only if $\theta_{-} \theta_{+} > 0$. In consequence, the disappearance of the trapped region at $v=v_d$ implies $\theta_{-} \theta_{+} \vert_{v_\mathrm{d}} \leqslant 0$. At this point, the inner and outer horizon merge, i.e.\ $r_-(v_\mathrm{d}) \equiv r_+(v_\mathrm{d})$. From Eqs.~\eqref{eq:genRBH.fvr} and \eqref{eq:rad.null.geod.exp}, it then follows that 
\begin{align}
	\theta_{-} \theta_{+} \big\vert_{v_\mathrm{d}} = -\frac{2}{r^2} g(v_\mathrm{d},r) \big(r-r_{+}(v_\mathrm{d})\big)^{a+b} \leqslant 0 .
\end{align}
For this to hold true $\forall r \geqslant 0$, we must have $g(v,r)>0$, and the sum of the powers $a+b$ must be even (otherwise the trapped region cannot disappear, which would imply that an RBH cannot evaporate completely). In addition, the sign of the outgoing radial null geodesic $\theta_+$ can only change (which is a prerequisite to enable the formation of a trapped region) if both $a$ and $b$ are odd numbers due to the fact that the immediate neighborhood of the RBH center must be an untrapped region \cite{ms:23}.

A model that --- once generalized to the dynamical case [i.e.\ $r_-' \neq 0$ and $r_+' \neq 0$] --- satisfies all of the requirements outlined above is the inner-extremal RBH model introduced in Ref.~\citenum{cdlpv:22} (whose associated mathematical expressions are indicated by the subscript ``ie'' in what follows). In the dynamical case, it is described by the metric function
\begin{align}
	f_{\text{ie}}(v,r) &= g_{\text{ie}}(v,r) (r-r_{-})^3 (r-r_{+}) , 
	\label{eq:ieRBH.fvr} \\
	g_{\text{ie}}(v,r) &= \frac{1}{r^3_{-}r_{+} - (3r^2_{-}r_{+} + r^3_{-}) r + c_2 r^2 - 3r_{-}r^3 + r^4} , 
	\label{eq:ieRBH.gvr}
\end{align}
where
\begin{align}
	c_2= \tilde{c}_2 + \frac{15}{4} r^2_{-} + \frac{9}{4} r_{-} r_{+} + \frac{1}{4} \frac{r^3_{-}}{r_{+}} , \quad \tilde{c}_2 \geqslant 0 ,
	\label{eq:ieRBH.c2coeff}
\end{align}
and we have omitted dependencies on $v$ [$r_\pm \equiv r_\pm(v)$, $c_2 \equiv c_2(v)$, $\tilde{c}_2 \equiv \tilde{c}_2(v)$] for the sake of readability. The explicit form of the coefficient $\tilde{c}_2$ is constrained by the positivity requirement of the MS mass. An explicit derivation of Eqs.~\eqref{eq:ieRBH.fvr}--\eqref{eq:ieRBH.c2coeff} including our choice for $\tilde{c}_2$ [cf.\ Eq.~\eqref{app:eq:ieRBH.coeffc2t.expl}] is provided in App.~\ref{app:ieRBH.fvr.gvr.deriv}. 

Since this is a dynamical model, we must consider appropriate generalizations of surface gravity to quasilocal non-Killing horizons \cite{ny:08,clv:13}. The two principal generalizations to dynamical black hole spacetimes are related to either the affine peeling surface gravity \cite{blsv:11} or the so-called Kodama surface gravity \cite{k:80,av:10,kpv:21}. Here, we restrict our considerations to the latter since the peeling surface gravity is ill-defined for transient objects that form in finite time of a distant observer \cite{mt:21,mt:23,mmt:22} (which includes dynamical RBHs), and there are strong arguments that Kodama surface gravity is the critical quantity with respect to the emission of Hawking radiation \cite{kpv:21,mmt:22}.

It is worth noting that any dynamical generalization of surface gravity vanishes at a degenerate horizon \cite{h:94}. Thus, the inner [$a=3$] and outer [$b=1$] horizon degenaracies of the inner-extremal RBH model [cf.\ Eqs.~\eqref{eq:genRBH.fvr} and \eqref{eq:ieRBH.fvr}] imply a nonzero Kodama surface gravity at the outer horizon and a vanishing Kodama surface gravity at the inner horizon. The latter is what ultimately cures the mass inflation instability problem that typically plagues RBHs \cite{cdlpv:22} (at least in classical GR).

The assumption that a regular (in the sense that scalar curvature invariants remain finite) apparent horizon $r_\sg$ forms in finite time of a distant observer implies a violation of the NEC \cite{mmt:rev:22}. This leaves only evaporating black holes [$r_\sg' < 0$] and accreting white holes [$r_\sg' > 0$] as viable (i.e.\ real-valued) horizonful solutions to the spherically symmetric semiclassical Einstein equations (see Ref.~\citenum{mmt:rev:22}, Table 2). As our interest lies in describing singularity-free objects resulting from gravitational collapse, we limit our considerations to evaporating RBH solutions [i.e.\ $r_-' < 0$ and $r_+' < 0$] in what follows. In this case, the inner and outer components of the apparent horizon are timelike membranes for the entire evolution $v \in (v_\mathrm{f},v_\mathrm{d})$ of the black hole, i.e.\ from immediately after the instant of horizon formation at $v=v_\mathrm{f}$ until just before the disappearance of the trapped region at $v=v_\mathrm{d}$ \cite{dst:22}. This can be seen as follows: a hypersurface $\Sigma$ can be defined by restricting the coordinates via $\Phi(\Sigma_{r_\xi}) \eqdef r - r_\xi \equiv 0$. The inner and outer horizon correspond to the constraint
\begin{align}
	\Phi(\Sigma_{r_\pm}) = r - r_\pm \equiv 0 ,
\end{align} 
which leads to a normal vector $n_{\mu}$ defined by
\begin{align}
	n_{\mu} \eqdef \eta \partial_\mu \Phi(\Sigma_{r_\pm}) = \eta \left(-r'_\pm, 1, 0, 0 \right) ,
\end{align} 
where $\eta$ is a normalization factor. Using the simplified metric of Eq.~\eqref{eq:simpl.metr}, the inner product of this normal vector at the inner/outer horizon is given by
\begin{align}
	n_\mu n^\mu \big\vert_{r_\pm} = - 2 \eta^2 r'_\pm .
\end{align}
For evaporating RBHs [$r_\pm' < 0$] this inner product is spacelike, $n_\mu n^\mu > 0$. Consequently, the causal character of the inner and outer horizon is timelike.

\section{Energy momentum tensor and energy conditions for an evaporating regular black hole} \label{sec:EMT.NEC}

\subsection{Hawking--Ellis classification} \label{subsec:EMTclass}
The Hawking--Ellis classification of the EMT \cite{he:book:73} provides a convenient model-independent framework to study the matter content of a spacetime geometry as well as its connection to various energy conditions \cite{mv:18,mv:21}. The different EMT types \RN{1}-\RN{4} are distinguished through their eigenvectors, and more specifically through their causal structure (timelike vs.\ null vs.\ spacelike) and degeneracy (single, double, triple). We work in an orthonormal frame (abbrev.\ ONF; indicated by a hat `` $_{\hat{}}$ '' on the tensor indices in what follows) in which the eigenvalues $\lambda$ of the EMT are the solutions of the equation 
\begin{align}
	\det \left( T_{\hat{\mu}\hat{\nu}} - \lambda \eta_{\hat{\mu}\hat{\nu}} \right) = 0 .
\end{align}
In spherical symmetry, the generic form of the EMT in the ONF is given by
\begin{align}
	T_{\hat{\mu}\hat{\nu}} =
	\scalebox{0.5}[1]{$\displaystyle
		\left(\scalebox{2}[1]{$\displaystyle
		\begin{array}{cccc} \\[-3mm]
			T_{\hat{0}\hat{0}} & T_{\hat{0}\hat{1}} & 0 & 0 \\
			T_{\hat{1}\hat{0}} & T_{\hat{1}\hat{1}} &0 & 0 \\
			0 & 0 & T_{\hat{2}\hat{2}} & 0\\
			0 & 0 & 0 & T_{\hat{3}\hat{3}}
			\vspace*{1mm}
		\end{array}$}
		\right)$} 
		\; ,
\end{align}
where explicit expressions for the EMT components are provided in App.~\ref{app:EMTcomp.ONF}, and $T_{\hat{2}\hat{2}}=T_{\hat{3}\hat{3}}$ due to the symmetries of the spacetime. Both of these components are also eigenvalues of the EMT. Their corresponding eigenvectors are $v_{\hat{2}}=(0,0,1,0)$ and $v_{\hat{3}}=(0,0,0,1)$, respectively. The remaining eigenvalues are determined by the (reduced) characteristic polynomial
\begin{align}
	\lambda^{2} + \left( T_{\hat{0}\hat{0}} - T_{\hat{1}\hat{1}} \right) \lambda + \left( T^{2}_{\hat{0}\hat{1}} - T_{\hat{1}\hat{1}} T_{\hat{0}\hat{0}} \right) = 0 ,
	\label{eq:EMT.EVs}
\end{align}
and their degeneracy is specified by its discriminant
\begin{align}
	\Delta = \left( T_{\hat{0}\hat{0}} + T_{\hat{1}\hat{1}} \right)^2 - 4 T^2_{\hat{0}\hat{1}} .
	\label{eq:discr}
\end{align} 
For the general spherically symmetric metric specified in Eq.~\eqref{eq:gen.metr}, the discriminant is given by 
\begin{align}
	\Delta(v,r) = \frac{\partial_{r} h_{+}}{16 \pi^2 r^2} \left[ (\partial_{r} h_{+}) f^2 - 2 e^{-h_{+}} \partial_{v} f \right] . 
	\label{eq:gen.discr}
\end{align}
For $h_{+}=0$ we have $\Delta=0$, meaning that  --- depending on the value of the EMT component $T_{\hat{0}\hat{1}}$ --- the EMT will be either of type \RN{2} or degenerate type \RN{1} \cite{mv:21}. Dynamical RBHs necessarily have $T_{\hat{0}\hat{1}} \neq 0$. In this case, there exists only one additional eigenvector, and the corresponding EMT is of type \RN{2} (null dust/massless radiation). For static RBH models, on the other hand, $T_{\hat{0}\hat{1}}=0$, $T_{\hat{1}\hat{1}}=-T_{\hat{0}\hat{0}}$, and the EMT admits four eigenvectors corresponding to type \RN{1}, which is degenerate in this case because two different eigenvectors are associated with the same eigenvalue. These results are consistent with the theorem derived in Ref.~\citenum{m:21}. The same type of EMT arises for the cosmological constant fluid \cite{mv:21}. However, in contrast to the perfect fluid associated with the cosmological constant, RBH models (both static and dynamical) correspond to an imperfect (anisotropic) fluid since $T_{\hat{1}\hat{1}} \neq T_{\hat{2}\hat{2}}$ \cite{rz:13,ac:21}.

The type of EMT that is associated with a particular solution of the Einstein equations has implications for the status of various energy conditions \cite{he:book:73}. Here, we focus on the null energy condition (NEC), which posits that $T_{\mu\nu} \ell^\mu \ell^\nu \geqslant 0$, i.e.\ the contraction of the EMT with any future-directed null vector $\ell^\mu$ is non-negative. Since the NEC is the weakest of all energy conditions, its violation implies that all other energy conditions (strong, weak, dominant) are violated as well. In spherical symmetry, an EMT of type \RN{4} is generally considered to be the most exotic as it always violates the NEC. The expectation value of the renormalized EMT near the apparent horizon of a spherically symmetric black hole has been shown to be of this type \cite{r:86,mv:13}. Conversely, as we have outlined above, dynamical RBHs are described by an EMT of type \RN{2}. This raises the question of how the NEC behaves in the vicinity of the inner and outer horizon, as well as inside of the RBH's trapped region.   

\subsection{Null energy condition} \label{subsec:NEC}
An apparent horizon can be observed from future null infinity $\mathscr{I}^{+}$ only if the NEC is violated in its vicinity \cite{he:book:73,bhl:18,bmmt:19}. For PBHs (which include RBHs as the singularity-free subset) with $h_+=0$, the NEC is marginally satisfied for ingoing null vectors \cite{mmt:rev:22}. For the metric specified in Eq.~\eqref{eq:simpl.metr}, an outgoing null vector is described by $\ell^\mu = (1,f/2,0,0)$. Since the NEC is a classical energy condition, it is reasonable to expect violations in the presence of quantum effects that ultimately lead to the evaporation of RBHs \cite{ks:20}. In fact, it has been demonstrated that the emission of Hawking radiation \cite{h:74,h:75} violates several energy conditions \cite{v:97,fn:98,w:01,lo:16}. It should also be noted that --- contrary to popular belief --- an event horizon is not required to enable the emission of Hawking radiation: a slowly evolving future apparent horizon is sufficient \cite{v:03}. 

Using the components of the Einstein tensor $G_{\mu\nu} = 8 \pi T_{\mu\nu}$, the NEC can be expressed as
\begin{align}
	T_{\mu\nu} \ell^\mu \ell^\nu = \frac{1}{8\pi} \left(G_{00} + G_{01} f + G_{11} \frac{f^2}{4}\right) \geqslant 0 ,
\end{align}
evaluated at the horizon. For the metric of Eq.~\eqref{eq:simpl.metr}, we find
\begin{align}
	T_{\mu\nu} \ell^\mu \ell^\nu = - \frac{\partial_v f}{8\pi r} . 
	\label{eq:NEC}
\end{align}
As argued in Sec.~\ref{sec:RBHs}, we consider only evaporating RBH solutions [$r_-'<0$ and $r_+'<0$]. Evaluating Eq.~\eqref{eq:NEC} at the horizons, we find that the NEC is violated in the vicinity of the outer horizon while being satisfied in the vicinity of the inner horizon. Assuming that semiclassical gravity is valid, this is a universal property for spherically symmetric dynamical RBHs. As we will see below, the change in the signature of the NEC expression implies the presence of a hypersurface $r_0(v)$ that acts as a boundary between the NEC-violating and the NEC-non-violating region within the trapped spacetime domain.

For the generic metric function Eq.~\eqref{eq:genRBH.fvr} describing dynamical RBHs, performing a series expansion of Eq.~\eqref{eq:NEC} at the outer apparent horizon yields
\begin{align}
	T_{\mu\nu} \ell^\mu \ell^\nu \big\vert_{r_+} 
	& = \frac{-b g(v,r_{+}) (-r'_{+}) (r_{+}-r_{-})^a}{8 \pi r_{+}} (r-r_{+})^{b-1} \nonumber \\
	& \hspace*{33mm} + \mathcal{O}{\left( r-r_{+} \right)^b} .
\end{align}
Therefore, the NEC is always violated in the vicinity of the outer horizon (both inside and outside of the trapped region) irrespective of its degeneracy, as our argumentation in Sec.~\ref{sec:RBHs} revealed that $b$ is odd. For a degenerate outer horizon [$b>1$], the NEC is marginally satisfied at the outer horizon $r=r_+$ itself. However, note that having a nonzero surface gravity at the outer horizon is only possible if it is nondegenerate \cite{h:94}, i.e.\ if $b=1$ [cf.\ Eq.~(12) in Ref.~\citenum{ms:23}]. Once again, we stress that the violation of the NEC is not only consistent with, but rather, it is a prerequisite for the observability of the outer apparent horizon from future null infinity. 

At the inner horizon, we find
\begin{align}
	T_{\mu\nu} \ell^\mu \ell^\nu \big\vert_{r_-} 
	& = (-1)^{b+1} \frac{a g(v,r_{-})(-r'_{-})}{8 \pi r_{-}} (r_{+}-r_{-})^{b} \nonumber \\
	& \qquad \cdot (r-r_{-})^{a-1} + \mathcal{O}{\left( r-r_{-} \right)^{a}} .
\end{align}
Consequently, the NEC is satisfied in the vicinity of the inner horizon, and marginally satisfied at $r=r_-$ if it is degenerate [$a>1$]. Since the NEC is always violated in the vicinity of the outer apparent horizon, consistency requires the existence of a hypersurface $r_{0} \in (r_{-},r_{+})$ located between the inner and outer horizon, thus effectively separating the trapped region into two distinct spacetime domains as detailed in Tab.~\ref{tab:NECregions}.

To illustrate this behavior in more detail, we consider the dynamical generalization of the inner-extremal RBH model \cite{cdlpv:22} described by Eqs.~\eqref{eq:ieRBH.fvr}--\eqref{eq:ieRBH.c2coeff}. Evaluating Eq.~\eqref{eq:NEC} at the outer horizon for the metric function of Eq.~\eqref{eq:ieRBH.fvr}, we find
\begin{align}
	T_{\mu\nu} \ell^\mu \ell^\nu \big\vert_{r_+} 
	& = \frac{(r_{+}-r_{-})^3 r'_{+}}{8 \pi r^{3}_{+} (r^2_{+} + r^{2}_{-})} ,
\end{align}
which is either negative or zero due to $r_+'<0$; equality occurs only at the extremal cases, i.e.\ at the formation $v=v_\mathrm{f}$ and disappearance $v=v_\mathrm{d}$ of the trapped region. At the inner horizon, the corresponding expression is always zero due to its degenerate nature [$a=3$ in the inner-extremal RBH model, cf.\ Eq.~\eqref{eq:ieRBH.fvr}], and thus the NEC is always satisfied. At the formation $v=v_\mathrm{f}$ and disappearance $v=v_\mathrm{d}$ of the trapped region, the two expressions for the NEC coincide as the inner and outer apparent horizon (e)merge.

\begin{figure}[!htbp]
	\includegraphics[scale=0.60]{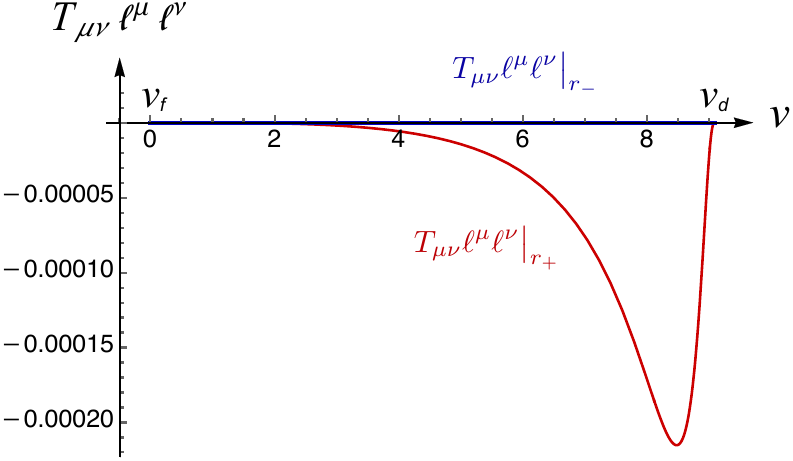}
	\caption{Evaluation of the NEC for the entire evolution period $v \in \left[ v_\mathrm{f} , v_\mathrm{d} \right]$ of an evaporating inner-extremal RBH. The red (blue) line represents the NEC evaluated at the outer (inner) apparent horizon $r_+(v)$ $\left[r_{-}(v)\right]$. The points $v=v_\mathrm{f}$ and $v=v_\mathrm{d}$ denote the advanced null coordinate at the formation and disappearance of the trapped region, respectively. For the purpose of modeling the NEC violation, the outer and inner horizon are chosen as $r_{+}(v) = r_{+}(0) - a_{+} v - b_{+} v^2$ and $r_{-}(v) = r_{-}(0) - a_{-} v - b_{-} v^2$ with parameter values $a_{+}=0.1$, $b_{+}=0.1$, $a_{-}=1$, $b_{-}=0.001$, and initial radii $r_{+}(0) = r_{-}(0) = 10$. With this particular choice, the equation $r_{+}(v)=r_{-}(v)$ has exactly two roots, which represent the formation and the disappearance point.}
	\label{fig:NEC}
\end{figure}

Fig.~\ref{fig:NEC} illustrates the extent of the NEC violation for an evaporating inner-extremal RBH. We note that the minimum (i.e.\ a maximally violated NEC) occurs close to the disappearance of the trapped region, indicating that the quantum effects responsible for the NEC violation are more pronounced towards the final stages of the evaporation process. This is in qualitative agreement with the results obtained for evaporating two-dimensional regular dilatonic black holes \cite{cos:23}.

The existence of an NEC boundary has been pointed out previously for RBHs with horizons that are not exclusively timelike \cite{bhl:18}. However, in our analysis, both the inner and outer horizon are timelike for the entire evolution of the evaporating RBH as motivated by our argumentation in Sec.~\ref{sec:RBHs}. Determining the location of the hypersurface separating the NEC-violating from the NEC-non-violating region requires solving the schematic equation
\begin{align}
	T_{\mu\nu} \ell^\mu \ell^\nu \stackrel{!}{=} 0 .
\end{align}
For the inner-extremal RBH model [that is, using Eq.~\eqref{eq:NEC} with the metric functions Eqs.~\eqref{eq:ieRBH.fvr}--\eqref{eq:ieRBH.c2coeff}], we obtain five roots, namely $r=0$, $r=r_{-}$ (double root), $r_0$, and two complex conjugate roots which are excluded since $r$ is real. The expression obtained for the root $r_0$ is too convoluted to determine its exact location. However, in the vicinity of the inner apparent horizon $r \sim r_-$, we find 
\begin{align}
	T_{\mu\nu} \ell^\mu \ell^\nu \big\vert_{r \sim r_-} 
	&= \frac{-3r'_{-} (r_{+}-r_{-})}{8 \pi r^4_{-} (r_{+}+r_{-})}(r - r_{-})^2 + \mathcal{O}{(r-r_{-})^3} ,
\end{align}
indicating that the NEC is satisfied as $r_-'<0$.
\begin{table}[!htpb] 
	\centering
	\begin{tabular}{ >{\raggedright\arraybackslash}m{0.20\linewidth}  @{\hskip 0.05\linewidth} >{\centering\arraybackslash}m{0.20\linewidth} @{\hskip 0.05\linewidth} >{\centering\arraybackslash}m{0.20\linewidth} @{\hskip 0.055\linewidth}  >{\centering\arraybackslash}m{0.20\linewidth}}
		& $0 \leqslant r \leqslant r_{-}$ & $r_{-} < r \leqslant r_{0}$ & $r_{0} < r \leqslant r_{+}$ \\ \toprule
		$T_{\mu\nu} \ell^\mu \ell^\nu \stackrel{?}{\geqslant} 0$ & $\cmark$ & $\cmark$ & $\xmark$ 
		\\[1mm] \bottomrule
	\end{tabular}
	\caption{Overview of NEC-non-violating ($\cmark$) and NEC-violating ($\xmark$) regions of an evaporating RBH with a nondegenarate outer horizon [$b=1$]. If the outer horizon is degenerate [$b>1$], the NEC-violating region is given by $r_{0} < r < r_{+}$, i.e.\ it no longer includes the outer horizon $r=r_+$ itself.}
	\label{tab:NECregions} 
\end{table}
\begin{figure*}[!htpb]
	\centering
	\begin{tabular}{@{\hspace*{0.006\linewidth}}p{0.45\linewidth}@{\hspace*{0.05\linewidth}}p{0.45\linewidth}@{}}
		\centering
		\subfigimg[scale=0.425]{(a)}{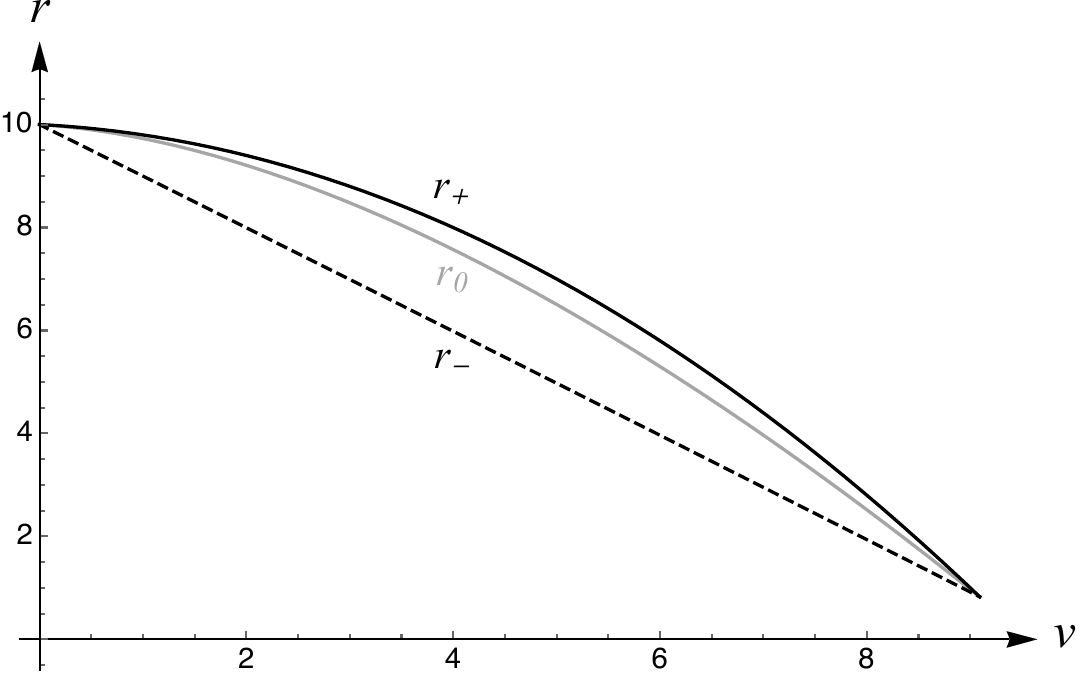} &
		\subfigimg[scale=0.40]{(b)}{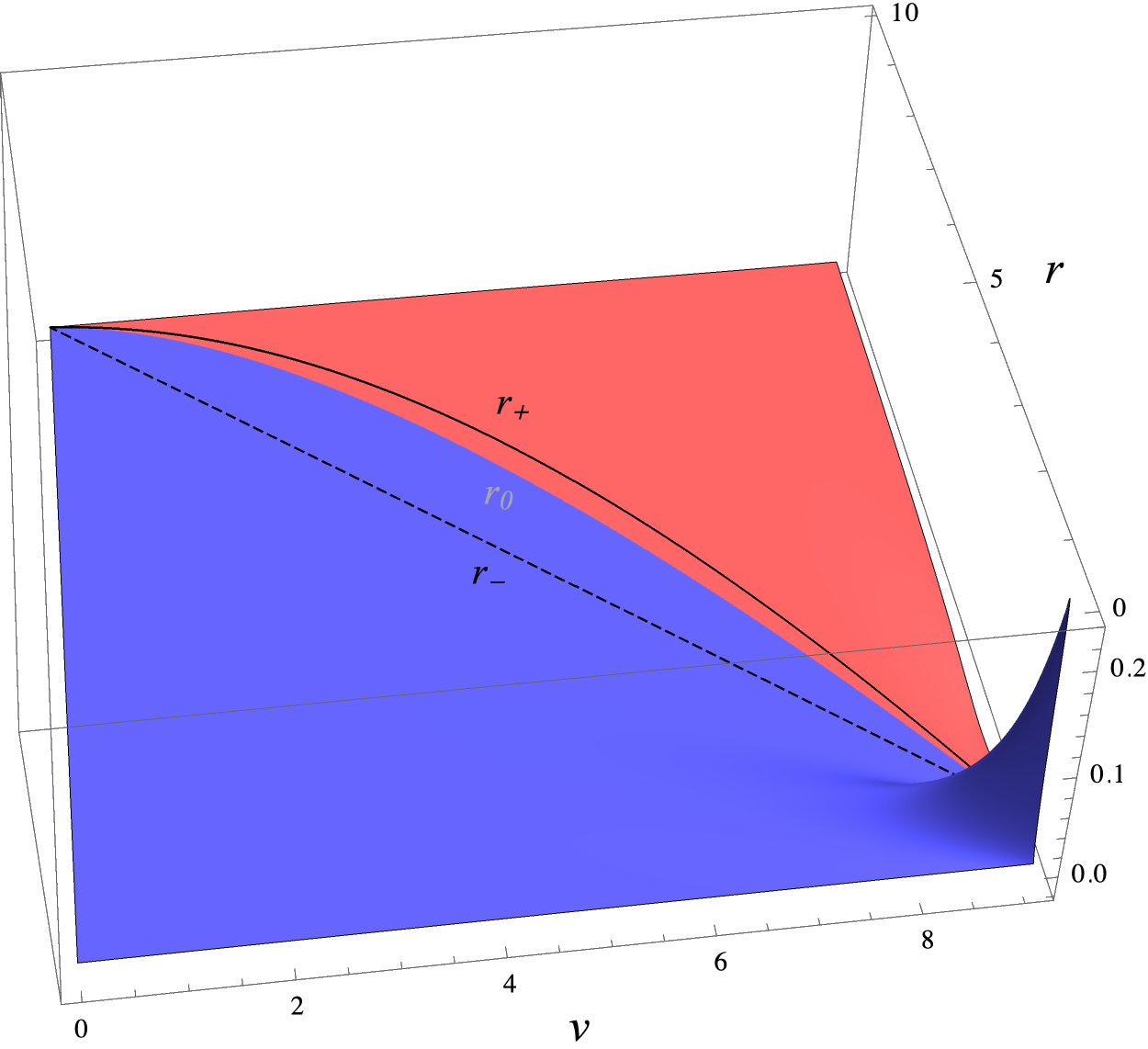}
	\end{tabular}
	\caption{The solid and dashed black lines represent the evolution of the outer $r_+$ and inner $r_-$ apparent horizon, respectively, until the disappearance of the trapped region. The solid gray line indicates the location of the hypersurface $r_0$ which separates the NEC-violating region between $r_0$ and $r_+$ from the NEC-non-violating region between $r_0$ and $r_-$. (a) Illustration of the trapped spacetime region in the $v$-$r$-plane. (b) Status of the NEC within the trapped region and its vicinity. The spacetime domain where the NEC is violated (satisfied) is shaded in red (blue). The hypersurface $r_{0}$ located between the two horizons corresponds to the boundary of the two domains.}
	\label{fig:NEC.3D}
\end{figure*}

Our analysis in this section is valid for $v_\mathrm{f} \leqslant v \leqslant v_\mathrm{d}$, i.e.\ for as long as the trapped spacetime region persists, including the instants of its formation and disappearance. Fig.~\ref{fig:NEC.3D} illustrates the trapped spacetime region and the status of the NEC over the course of the RBH's lifetime. In the next section, we consider the energy density observed by a moving observer, and find that a similar feature of the trapped region emerges.

\section{Trajectories of moving observers} \label{sec:particle.motion}
In this section, we study the trajectory of a moving observer Alice. We start by examining the energy density she observes while following a particular geodesic that allows her to enter and exit the trapped region.

\subsection{Ingoing and outgoing trajectories:  \\ Entering and exiting the trapped region}
Assume that Alice begins her expedition from the untrapped region near the RBH center and initially follows an outgoing geodesic. After making a few generic statements about her trajectory, we once again illustrate our results based on the dynamical inner-extremal model described by Eqs.~\eqref{eq:ieRBH.fvr}--\eqref{eq:ieRBH.c2coeff} and comment on important features that arise over the course of her journey.

For radially moving timelike observers and particles, the Lagrangian associated with the metric of Eq.~\eqref{eq:simpl.metr} is given by 
\begin{align}
	\mathcal{L} = \frac{1}{2} f \dot{v}^2 - \dot{v} \dot{r} ,
\end{align}
where the overdot denotes derivatives with respect to Alice's proper time\footnote{Since the action and its corresponding Lagrangian are invariant under reparametrization of the trajectory and our argumentation in this section requires only the equations of motion, we have the freedom to choose the time parameter in the Lagrangian.}. The corresponding Euler--Lagrange equations are
\begin{align}
	\ddot{v} &= -\frac{1}{2}(\partial_{r}f)\dot{v}^2 \label{eq:ddv} , \\
	\ddot{r} &= \frac{1}{2}(\partial_v f)\dot{v}^2-\frac{1}{2}(\partial_rf) \label{eq:EL.ddr} .
\end{align} 
Alice's four-velocity is normalized by $u^{\mu}u_{\mu}=-1$, which results in
\begin{align}
	-f \dot{v}^2 + 2 \dot{v} \dot{r} = - 1 .
\end{align}
From the normalization conditions, we obtain two possible solutions for $\dot{v}$, which in turn restrict the admissible values of the radial velocity $\dot{r}$. Specifically, we find that
\begin{align}	
	\dot{v} = \frac{\dot{r} \pm \sqrt{\dot{r}^2+f}}{f} . \label{eq:dv.pm}
\end{align}
To ensure that Eq.~\eqref{eq:dv.pm} maintains real-valued solutions inside of the trapped region, the relation
\begin{align}
	\dot{r} \leqslant -\sqrt{-f}
\end{align}
must hold $\forall v \in \left(v_\mathrm{f},v_\mathrm{d}\right)$. Outside of the trapped region there are no restrictions since $f(v,r)$ is always positive there. It is important to note that, during Alice's motion, $\dot{v}$ must be positive. This condition necessitates different choices for the sign in the numerator of Eq.~\eqref{eq:dv.pm} to accommodate different types of motion associated with the spacetime regions Alice is traversing. For untrapped regions, i.e.\ for $0 \leqslant r < r_{-}$ and $r > r_{+}$, it is straightforward to verify that the correct choice of signature is ``$+$" for both ingoing and outgoing trajectories. On the other hand, this is not the case when trajectories inside the trapped region are considered. The deviation occurs because $\dot{r}<0$ for both ingoing and outgoing trajectories, which is a well-known feature associated with the presence of a trapped region due to the fact that $f(v,r)$ is negative there \cite{mtw:book:73,dst:22,b:lect:23}. In this case, to guarantee the positivity of $\dot{v}$ and ensure that it takes on only real values, ingoing and outgoing trajectories must satisfy the respective relations
\begin{align}
	\dot{v} &= \frac{\dot{r}+ \sqrt{\dot{r}^2+f}}{f}, \quad r_{-}<r<r_{+} \quad \text{(ingoing)} \; ,
	\label{eq:dv.in} \\
	\dot{v} &= \frac{\dot{r}- \sqrt{\dot{r}^2+f}}{f}, \quad r_{-}<r<r_{+} \quad \text{(outgoing)} \; .
	\label{eq:dv.out}
\end{align}
With this in mind, an interesting exercise is to investigate what happens to the radial acceleration of Alice's motion in the general case described by Eq.~\eqref{eq:genRBH.fvr}. As Alice begins her journey from an untrapped region, her trajectory (both ingoing and outgoing) is described by the signature given in Eq.~\eqref{eq:dv.in}. On her outgoing trajectory, she first encounters the inner horizon. Performing a series expansion on the RHS of Eq.~\eqref{eq:EL.ddr} reveals that, when approaching the inner horizon on an outgoing geodesic that satisfies Eq.~\eqref{eq:dv.in}, her radial acceleration is given by
\begin{align}
	\ddot{r} = \frac{(-1)^{a+b} 2a \dot{r}^2}{g(v,r_{-}) (r_{+}-r_{-})^{b}} \frac{r'_{-}(v)}{|r-r_{-}|^{a+1}} + \mathcal{O}{\left( \frac{1}{(r-r_{-})^{a}} \right)} . 
\end{align}
From our argumentation in Sec.~\ref{sec:RBHs}, we know that
\begin{enumerate}[label=\Alph*.]
	\item the disappearance of a trapped region in finite time according to the clock of a distant observer is only possible if the sum of the horizon degeneracies $a+b$ is an even number;
	\item the only viable dynamically evolving semiclassical black hole solutions in spherical symmetry are evaporating black holes; for dynamical RBHs, this implies that both the inner and outer component of the apparent horizon must shrink.
\end{enumerate}
Therefore, while traveling on an outgoing trajectory, Alice's acceleration is negative and divergent, thus forcing her to stop and reverse her trajectory so as to follow an ingoing geodesic. We note that the value $\dot{r}=0$ for the radial velocity is allowed since Alice is still in the untrapped region $r<r_{-}$. Assuming that semiclassical gravity remains valid in this regime, this is another universal result that holds for every dynamical RBH described by Eq.~\eqref{eq:genRBH.fvr}. 

\begin{figure*}[!htbp]
	\centering
	\begin{tabular}{@{\hspace*{0.02\linewidth}}p{0.45\linewidth}@{\hspace*{0.05\linewidth}}p{0.45\linewidth}@{}}
		\centering
		\subfigimg[scale=0.40]{(a)}{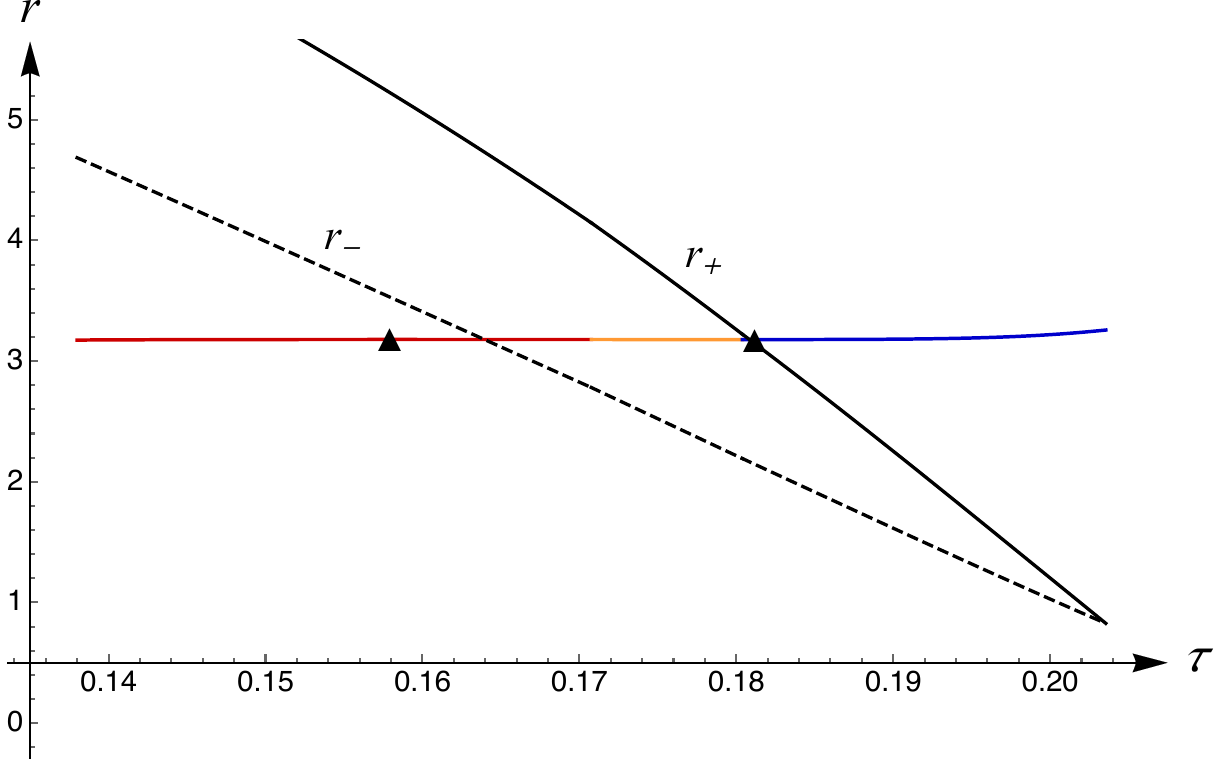} &
		\subfigimg[scale=0.40]{(b)}{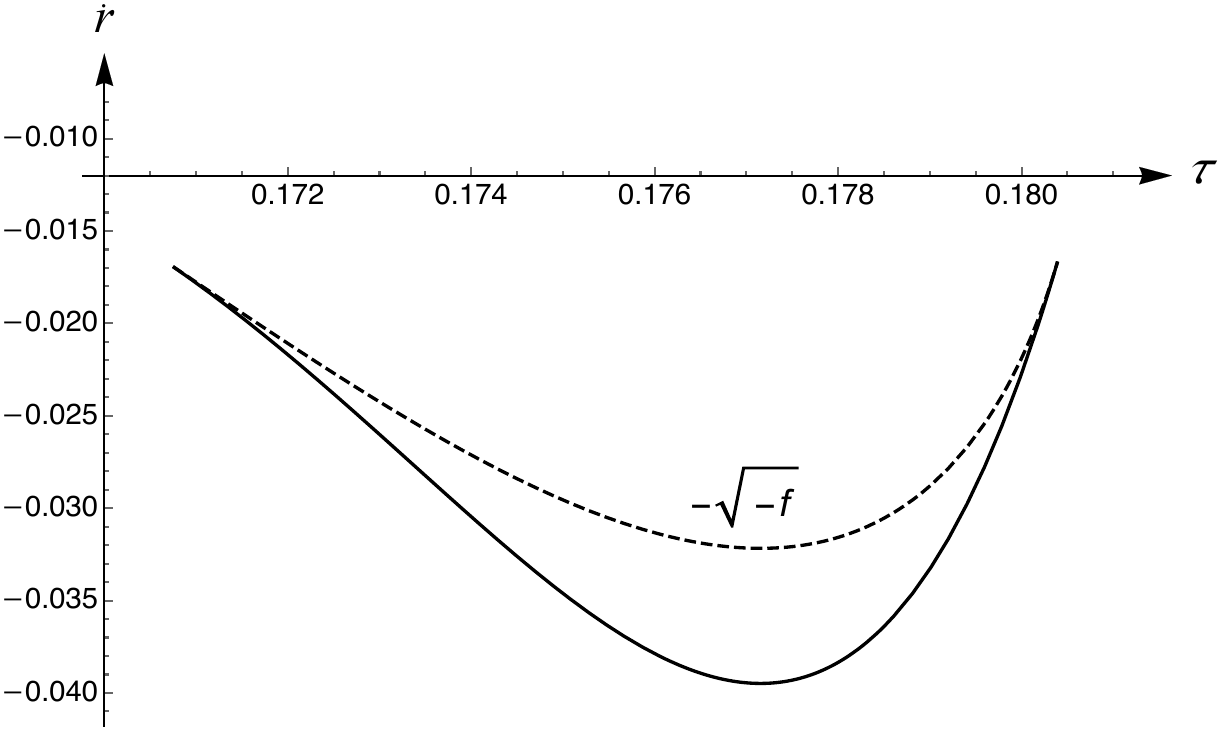}
	\end{tabular}
	\caption{Properties of outgoing trajectories illustrated using the generalized dynamical inner-extremal RBH model described by Eqs.~\eqref{eq:ieRBH.fvr}--\eqref{eq:ieRBH.c2coeff}. (a) An observer Alice starts her journey at areal radius $r<r_{-}$ and follows an outgoing trajectory. The red line represents the outgoing motion up to the first black triangle, which signifies the point where $\dot{r}=0$. The second segment of the red line corresponds to an ingoing trajectory. The orange line represents Alice's outgoing trajectory inside of the trapped region. The first segment of the blue line corresponds to an ingoing geodesic up to the black triangle, and the second segment describes an outgoing geodesic. (b) The solid black line illustrates the evolution of Alice's radial velocity $\dot{r}$ as a function of her proper time, and the dashed line the evolution of $-\sqrt{-f}$ during the outgoing motion of Alice inside of the trapped region (indicated by the orange segment in (a)). We note that the beginning and end of Alice's outgoing trajectory correspond to the points where her radial velocity equals the maximum possible value $\dot{r} = -\sqrt{-f}$.}
	\label{fig:trajectory}
\end{figure*}

Close to the evaporating inner horizon, specific trajectories undergo a transition from outgoing to ingoing. As a result, the inner horizon can overtake Alice, and she may find herself inside of the trapped region. In fact, this is the only way to enter the trapped region from areal radii $r<r_{-}$. Once inside, Alice's motion is characterized by $\dot{r}<0$ and $\ddot{r}<0$. This implies that her radial velocity $\dot{r}$ will continue to become increasingly more negative until it reaches the minimum value $-\sqrt{-f}$. At this point, a continuous transition from Eq.~\eqref{eq:dv.in} to Eq.~\eqref{eq:dv.out} occurs, and Alice will once again find herself on an outgoing geodesic. As she approaches the outer apparent horizon, her acceleration is given by 
\begin{align}
	\ddot{r} = \frac{2b \dot{r}^2}{g(v,r_{+}) (r_{+}-r_{-})^{a}} \frac{-r'_{+}(v)}{|r-r_{+}|^{b+1}} + \mathcal{O}{\left( \frac{1}{|r-r_{+}|^b} \right)} ,
\end{align}	
which is positive. Therefore, as she approaches the horizon, her velocity becomes less and less negative until she reaches the maximum allowed value. Once again, a transition occurs, this time from Eq.~\eqref{eq:dv.out} to Eq.~\eqref{eq:dv.in}, meaning that Alice now finds herself on an ingoing trajectory, and she can only exit the trapped region if the outer apparent horizon overtakes her. Consequently, no matter which horizon Alice approaches on an outgoing geodesic, there is a unique way to cross it, namely on an ingoing geodesic. This analysis provides relevant insights on how information (e.g.\ in the form of particles) can exit the trapped region, thus offering a natural resolution to the information loss problem \cite{h:16} as information that is supposedly trapped can once again become visible to an observer who is performing measurements in the exterior of the trapped region. However, as illustrated in Fig.~\ref{fig:trajectory}(a), information will become visible for distant observers only after the two horizons have merged and the trapped region has disappeared.

\subsection{Energy density at the horizon crossing}
In this subsection, we calculate the energy density experienced by Alice upon her exit from the trapped region. Recall that Alice's motion is described by the four-velocity
\begin{align}
	u^{\mu}_A = \left( \dot{v}, \dot{r}, 0, 0 \right) ,
	\label{eq:A.four-vel} 
\end{align}
and the energy density she observes is given by the contraction
\begin{align}
	\rho_A = T_{\mu\nu} u^{\mu}_A u^{\nu}_A ,
	\label{eq:A.rho}
\end{align}
or, equivalently,
\begin{align}
	\rho_{A} = \frac{1}{8\pi} \left(G_{00} \dot{v}^2 + 2 G_{01} \dot{v} \dot{r} + G_{11} \dot{r}^2 \right) .
	\label{eq:A.rho.Gten}
\end{align}
Instants where Alice crosses the horizons are of particular interest. Horizon crossings always occur while she is on an ingoing geodesic trajectory, meaning that $\dot{v}$ is given by Eq.~\eqref{eq:dv.in}, and $\dot{r}$ is negative at each crossing. Close to both horizons $f(v,r) \simeq 0$, and thus we can expand her ingoing trajectory [Eq.~\eqref{eq:dv.in}] and measured energy density [Eq.~\eqref{eq:A.rho.Gten}] as follows:
\begin{align}
	\dot{v} \big\vert_{r_\pm} &= - \frac{1}{2 \dot{r}} + \frac{f}{8 \dot{r}^3} + \mathcal{O}{\left( f^2 \right)} ,
	\label{eq:dv.ser} \\
	\rho_{A} \big\vert_{r_\pm} &= \left[ \frac{1}{8 \pi r^2} \left( 1 - r \partial_{r} f - \frac{r}{4 \dot{r}^2} \partial_{v} f \right) \right] \bigg \vert_{r_{\pm}} \hspace*{-1.25mm} + \mathcal{O}{(f)} .
	\label{eq:A.rho.ser}
\end{align}
Using Eqs.~\eqref{eq:dv.ser} and \eqref{eq:A.rho.ser}, we can confirm that Alice observes a finite energy density when crossing the horizons, although the specific values will depend on their degeneracy, see Eq.~\eqref{eq:A.rho.rm.nondeg}--\eqref{eq:A.rho.rp.deg}. Therefore, unlike non-geodesic observers, Alice does not experience any firewalls (in the sense of a diverging energy density) on her ingoing geodesic. When she is traveling in the vicinity of a nondegenerate inner horizon [$a=1$], we find that
\begin{align}
	\begin{aligned}
		\rho_{A} \big\vert_{r_{-}}^{\text{nondeg}} &= \frac{4 \dot{r}^2 - g(v,r_{-}) r_{-} (r_{-}-r_{+})^b (4 \dot{r}^2 - r'_{-})}{32 \pi \dot{r}^2 r^2_{-}} \\
		& \hspace*{42.5mm} + \mathcal{O}{(r-r_{-})} ,
	\end{aligned}
	\label{eq:A.rho.rm.nondeg}
\end{align}
whereas for a degenerate inner horizon [$a>1$] 
\begin{align}
	\rho_{A} \big\vert_{r_{-}}^{\text{deg}}  = \frac{1}{8\pi r^2_{-}} - \frac{r-r_{-}}{8 \pi r^3_{-}} + \mathcal{O}{(r-r_{-})^2} ,
	\label{eq:A.rho.rm.deg}
\end{align}
which is independent of the outer horizon degeneracy $b$. Similarly, we can calculate the relevant expressions when Alice travels near the vicinity of the outer horizon. For the nondegenerate case [$b=1$], we obtain 
\begin{align}
	\begin{aligned}
		\rho_{A} \big\vert_{r_{+}}^{\text{nondeg}} &= \frac{4 \dot{r}^2 - g(v,r_{+}) r_{+} (r_{+}-r_{-})^a (4 \dot{r}^2 - r'_{+})}{32 \pi \dot{r}^2 r^2_{+}} \\
		& \hspace*{42.5mm} + \mathcal{O}{(r-r_{+})} , 
	\end{aligned}
	\label{eq:A.rho.rp.nondeg}
\end{align}
while in the degenerate case $[b>1]$
\begin{align}
	\rho_{A} \big\vert_{r_{+}}^{\text{deg}} &= \frac{1}{8 \pi r^2_{+}} - \frac{r-r_{+}}{8 \pi r^3_{+}} + \mathcal{O}{(r-r_{+})^2} .
	\label{eq:A.rho.rp.deg}
\end{align}
Again, we demonstrate the physical implications of these results based on the dynamical inner-extremal model described by Eqs.~\eqref{eq:ieRBH.fvr}--\eqref{eq:ieRBH.c2coeff}. From Eq.~\eqref{eq:A.rho.ser}, or, equivalently, from Eqs.~\eqref{eq:A.rho.rm.deg} and \eqref{eq:A.rho.rp.nondeg} [recall that $a=3$ and $b=1$ in this model], we find that the energy density Alice observes at the degenerate inner horizon is given by 
\begin{align}
	\rho_A \big\vert_{r_{-}}^{a=3} = \frac{1}{8 \pi r^2_{-}} ,
	\label{eq:A.rho.ieRBH.rm}
\end{align}
which is positive, while at the nondegenerate outer horizon we obtain the more intricate expression
\begin{align}
	\rho_A \big\vert_{r_{+}}^{b=1} = \frac{4 \dot{r}^2 r_{-} (r^2_{-} - 2r_{-} r_{+} + 3 r^2_{+}) + (r_{+}-r_{-})^3 r'_{+}}{32 \pi \dot{r}^2 r^3_{+} (r^2_{+}+r^2_{-})} .
	\label{eq:A.rho.ieRBH.rp}
\end{align} 	
	
\begin{figure}[!htbp]
	\includegraphics[scale=0.425]{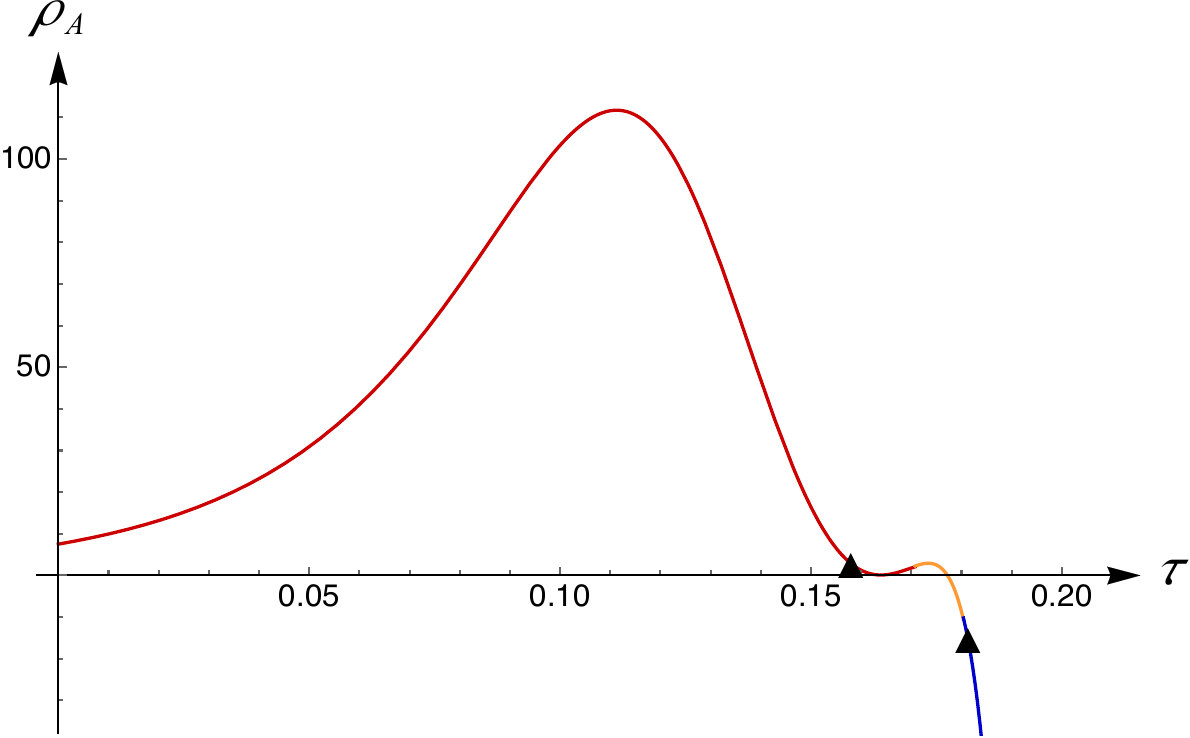}
	\caption{Energy density measured by a moving observer Alice throughout her trajectory as a function of her proper time. The red, orange, and blue lines as well as the two points indicated by black triangles have the same physical meaning as in Fig.~\ref{fig:trajectory}(a).} 
	\label{fig:Edens}
\end{figure}
Fig.~\ref{fig:Edens} represents the energy density Alice records throughout her journey. A negative energy density is observed close to the outer apparent horizon in agreement with the presence of an NEC-violating region, cf.\ Tab.~\ref{tab:NECregions}. In consonance with the results obtained in Subsec.~\ref{subsec:NEC}, we conclude that --- in the evolution of evaporating RBHs --- quantum effects are more pronounced near the outer apparent horizon and towards the final stages of the evaporation process. The absence of firewalls, i.e.\ the fact that the energy density observed by Alice upon crossing the horizons on her geodesic trajectory is always finite and never divergent, is a universal result due to the uniqueness of the horizon crossing scenario (there is no other way of crossing the horizon). Lastly, we note that at the disappearance point $v=v_\mathrm{d}$ of the trapped region [$r_-(v_\mathrm{d}) \equiv r_+(v_\mathrm{d})$], the expressions derived in Eqs.~\eqref{eq:A.rho.ieRBH.rm} and Eq.~\eqref{eq:A.rho.ieRBH.rp} coincide. While this is of course expected from self-consistency requirements, it can also be understood as an observable consequence of the transition from $k=0$ to $k=1$ PBH solutions, with the energy density taking on its appropriate value for the extremal case \cite{mmt:rev:22}. This will be explained further in the next section. 

\section{Transition from $k=0$ to $k=1$} \label{sec:transition}
With the assumptions of regularity and finite-time horizon formation (as described at the end of Sec.~\ref{sec:RBHs}), the semiclassical Einstein equations admit only two distinct classes of dynamical spherically symmetric black hole solutions, which are distinguished by the scaling behavior $\sim f^k$ [cf.\ Eq.~\eqref{eq:fvr.MSmass}] of their effective EMT components close to the horizon. The only self-consistent values are $k \in \lbrace 0, 1 \rbrace$ \cite{t:20,mt:21}. Hence, the two classes are typically referred to as $k=0$ and $k=1$ solutions. In addition to evaporating black holes [$r_\sg'<0$], each class also includes accreting white hole solutions [$r_\sg'>0$]. A detailed exposition of the two classes of solutions is provided in Chapter 2 of Ref.~\citenum{mmt:rev:22}, and a brief overview is given in Table \RN{1} of Ref.~\citenum{m:22}. Explicit relations for $k=0$ solutions in different coordinate systems can be found in \cite{dsst:23}. 

A straightforward way to determine which class of solutions a particular metric belongs to is the value of the linear coefficient $w_1(v)$ of its corresponding MS mass [cf.\ Eq.~\eqref{eq:MSmass}]: metrics with $w_{1}(v)<1$ belong to the $k=0$ class, whereas metrics in the $k=1$ class have $w_{1}(v)=1$. 

Rearranging Eq.~\eqref{eq:fvr.MSmass}, we can write
\begin{align}
	C(v,r) = r \big( 1-f(v,r) \big) .
	\label{eq:MSmass.fvr}
\end{align}  
By comparison with the series expansion of Eq.~\eqref{eq:MSmass}, we note that $w_1(v) = \partial_r C \vert_{r_+}$. Using Eq.~\eqref{eq:MSmass.fvr} in combination with the metric function Eq.~\eqref{eq:genRBH.fvr} describing generic dynamical RBHs, it is easy to confirm that $w_1(v)=1 \; \forall v \in \left[ v_\mathrm{f} , v_\mathrm{d} \right]$ when the outer horizon is degenerate [$b>1$]. On the other hand, if the outer horizon is nondegenerate [$b=1$], the linear coefficient of the MS mass is given by
\begin{align}
	w_{1}(v) \big\vert_{r_+} = 1 - g(v,r_{+}) r_{+} (r_{+}-r_{-})^{a} < 1,
	\label{eq:w1}
\end{align}
which is strictly smaller than one $\forall v \in (v_\mathrm{f},v_\mathrm{d})$ since $r_+ > r_- > 0$ by construction and $g(v,r)$ is positive (see Sec.~\ref{sec:RBHs}). Recall that at the instants of formation $v=v_\mathrm{f}$ and disappearance $v=v_\mathrm{d}$ of the trapped region, the inner and outer horizon (e)merge, i.e.\ $r_-(v_{\mathrm{f} \vert \mathrm{d}}) \equiv r_+(v_{\mathrm{f}|\mathrm{d}})$. From Eq.~\eqref{eq:w1}, we see that in this case $w_1(v_{\mathrm{f}|\mathrm{d}})=1$. In this sense, the value of the linear coefficient of the MS mass $w_1(v)$ indicates the transition from a $k=1$ to a $k=0$ solution at the formation of the trapped region [$w_1(v_\mathrm{f})=1 \to w_1(v>v_\mathrm{f})<1$], and similarly the transition from a $k=0$ to a $k=1$ solution at the disappearance of the trapped region [$w_1(v<v_\mathrm{d})<1 \to w_1(v_\mathrm{d})=1$].

\begin{figure}[!htbp]
	\includegraphics[scale=0.55]{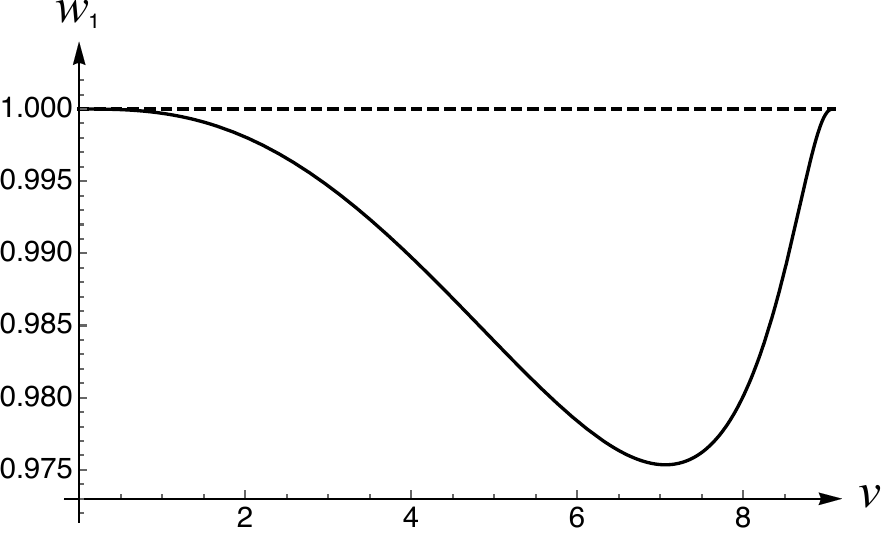}
	\caption{Evolution of the linear coefficient of the MS mass $w_1(v)$ from the instant of formation of the trapped region at $v_\mathrm{f}=0$ until its complete evaporation at $v_\mathrm{d} \approx 8.8$.} 
	\label{fig:w1}
\end{figure}

Fig.~\ref{fig:w1} illustrates the evolution of $w_1(v)$ for the dynamical inner-extremal RBH model described by Eqs.~\eqref{eq:ieRBH.fvr}--\eqref{eq:ieRBH.c2coeff}.
During the evaporation of the RBH, the value of $w_1(v)$ initially starts to decrease from its initial value $w_1(v_\mathrm{f})=1$ at the instant of formation. At some point in the evaporation process, it reaches a minimum value $w_1(v_m) \approx 0.975$ at $v_m \approx 7.15$. It then increases until the complete evaporation of the RBH at $w_1(v_\mathrm{d})=1$.

Apart from its role in modeling the transition between the two classes of dynamical PBH solutions in spherical symmetry, the linear coefficient of the MS mass $w_1(v)$ is of immediate relevance in the thermodynamic description of RBHs. In fact, it fully specifies the relevant work terms in the generalized dynamical first law of black hole mechanics [see Eq.~(22) in Ref.~\citenum{ms:23}]. A more comprehensive discussion of its role in black hole thermodynamics and specifically the first law of black hole mechanics is provided in Ref.~\citenum{ms:23}. 

To illustrate the connection with our results from Sec.~\ref{sec:particle.motion}, we note that the Kodama surface gravity at the horizon is given by
\begin{align}
	\kappa_K \big\vert_{r_\pm} = \frac{1}{2} \partial_r f \big\vert_{r_\pm} ,
\end{align}
which depends on the linear coefficient $w_1(v)$ of the MS mass due to the relations prescribed by Eqs.~\eqref{eq:fvr.MSmass} and \eqref{eq:MSmass}. In conjunction with Eq.~\eqref{eq:NEC}, this allows us to rewrite the energy density measured by a moving observer [cf.\ Eq.~\eqref{eq:A.rho.ser}] as
\begin{align}
	\rho_{A} \big\vert_{r_\pm} = \left( \frac{1}{8 \pi r^2} - \frac{\kappa_K}{4 \pi r} + \frac{T_{\mu\nu} \ell^\mu \ell^\nu}{4 \dot{r}^2} \right) \bigg\vert_{r_\pm} \hspace*{-0.55mm} + \mathcal{O}{(f)} .
\end{align}
At the outer horizon, the Kodama surface gravity is given by
\begin{align}
	\kappa_K \big\vert_{r_+} &= \frac{1-w_1(v)}{2r_+(v)} .
	\label{eq:Kodama.sg}
\end{align}
Due to the transition to a $k=1$ solution at the disappearance point $v=v_\mathrm{d}$ with $w_{1}(v_\mathrm{d})=1$, the Kodama surface gravity is zero when the evaporation process terminates. If we interpret the Kodama surface gravity as an effective temperature parameter, as is standard practice in black hole thermodynamics \cite{h:98}, this implies that the remnant is a cold remnant with zero temperature. Our analysis therefore confirms the previously obtained result that RBHs end their life as cold remnants \cite{cos:22} and demonstrates directly observable (at least in principle) physical consequences.

\section{Conclusions} \label{sec:concl}
Starting from the general metric function that describes generic spherically symmetric dynamical RBHs in semiclassical gravity [Sec.~\ref{sec:RBHs}, Eq.~\eqref{eq:genRBH.fvr}; see also Fig.~\ref{fig:RBH}] we derive their kinematic properties and analyze the behavior of the NEC in the vicinity of their horizons. Specifically, we find that the NEC is violated in the vicinity of the outer horizon and satisfied in the vicinity of the inner horizon, which implies that the trapped spacetime region (as determined from the behavior of null geodesic congruences) is effectively separated into an NEC-violating and an NEC-non-violating domain by a hypersurface situated between the two horizons [Subsec.~\ref{subsec:NEC}, Tab.~\ref{tab:NECregions} and Fig.~\ref{fig:NEC.3D}].

In addition, we demonstrate that there is a unique way for massive observers and particles to escape the trapped region on a geodesic trajectory, whereby crossing the horizon is only possible on an ingoing trajectory. This result has two rather pleasant side effects, namely i) the absence of firewalls (energy densities measured by the geodesic observer do not diverge); and ii) by virtue of particles (and thus any information content associated with their existence on the manifold) being able to escape the trapped spacetime region, a natural resolution to the information loss paradox [Sec.~\ref{sec:particle.motion}, Figs.~\ref{fig:trajectory} and \ref{fig:Edens}].

The presence of spacetime regions where the NEC is violated and the energy density is negative is inherently linked to the quantum effects underlying the evaporation of black holes. Our results suggest that, at least for dynamical RBHs, these effects are more dominant near the outer horizon and become more pronounced towards the final stages of the evaporation process [Sec.~\ref{sec:EMT.NEC}, Fig.~\ref{fig:NEC}]. 

The formation and disappearance of the trapped spacetime region are associated with a transition between the only two self-consistent classes of dynamical semiclassical black hole solutions in spherical symmetry. We show that the value of the linear coefficient of the MS mass can be used to distinguish the two classes of solutions and model the transition, and comment on implications for the thermodynamic description of RBHs [Sec.~\ref{sec:transition}, Fig.~\ref{fig:w1}]. 

While keeping our analysis as generic as possible, we highlight the importance of the degeneracy of the inner horizon (e.g.\ in curing mass inflation instabilities) and nondegeneracy of the outer horizon (e.g.\ in ensuring a nonzero Kodama surface gravity) where it is physically relevant. Throughout the article, we illustrate relevant features based on the dynamical generalization of the inner-extremal RBH model [Sec.~\ref{sec:RBHs}, Eqs.~\eqref{eq:ieRBH.fvr}--\eqref{eq:ieRBH.c2coeff}; see also Ref.~\citenum{cdlpv:22}]. Although our considerations here focus on spherically symmetric dynamical RBHs, realistic dark UCOs possess angular momentum, and modeling the emission of gravitational waves requires at least a (mass) quadrupole moment that cannot be captured in spherically symmetric settings. On this account, it will be interesting to extend the present analysis to nonsingular axially symmetric black hole spacetimes to confirm whether or not they exhibit similar physical features, and investigate what (if anything) can be learned about nonsingular black holes from the observed emission spectra of gravitational waves (e.g.\ the extraction of upper bounds on the minimal length scale $l$).

\section*{Acknowledgements}
We would like to thank Valentin Boyanov, Daniel Grumiller, Yasha Neiman, and Daniel Terno for helpful comments. SM is supported by the Quantum Gravity Unit of the Okinawa Institute of Science and Technology (OIST). IS is supported by an International Macquarie University Research Excellence Scholarship (IMQRES).
  
\appendix

\section{Derivation of the coefficients in the metric function of the inner-extremal regular black hole model  [Eqs.~\eqref{eq:ieRBH.fvr}--\eqref{eq:ieRBH.c2coeff}]} \label{app:ieRBH.fvr.gvr.deriv}
Here, we explicitly (re-)derive the metric functions Eqs.~\eqref{eq:ieRBH.fvr}--\eqref{eq:ieRBH.c2coeff} of the dynamical generalization of the inner-extremal RBH model \cite{cdlpv:22} and elaborate on the physical intuition for the underlying assumptions. Our starting point is Eq.~(2.9) of Ref.~\citenum{cdlpv:22}, which specifies the inner [$a=3$] and outer horizon [$b=1$] degenaracies, i.e.\ [cf.\ Eq.~\eqref{eq:genRBH.fvr}]
\begin{align}
	f(v,r) = g(v,r) (r-r_{-})^3 (r-r_{+}) ,
	\label{app:eq:ieRBH.fvr}
\end{align}
as well as the polynomial decomposition of $g(v,r)$ as a function of the radial coordinate $r$, namely
\begin{align}
	g(v,r) = \frac{1}{c_0 + c_1 r + c_2 r^2 + c_3 r^3 + c_4 r^4} , 
	\label{app:eq:ieRBH.gen.gvr}
\end{align} 
where explicit dependencies on the advanced null coordinate $v$ have been omitted here and in what follows unless they add mathematically relevant context. The more explicit form given in Eq.~\eqref{eq:ieRBH.gvr} is obtained by demanding that $g$ satisfy several requirements: first, we note that it is usually assumed that $f$ and $1/g$ are polynomials of the same degree in $r$ \cite{f:16}. Otherwise it would be impossible to recover the Vaidya form of the metric $f=1-r_+/r$ in the asymptotic limit $r\to\infty$. With respect to Eq.~\eqref{app:eq:ieRBH.gen.gvr}, the choice $a=3$ and $b=1$ then implies that higher-order powers $r>4$ in the denominator of $g$ are not permissible. Substituting Eq.~\eqref{app:eq:ieRBH.gen.gvr} into Eq.~\eqref{app:eq:ieRBH.fvr} and dividing by $r^4$, we obtain
\begin{align}
	f(v,r) = \frac{\left( 1 - \frac{r_{-}}{r} \right)^3 \left( 1 - \frac{r_{+}}{r} \right)}{\frac{c_0}{r^4} + \frac{c_1}{r^3} + \frac{c_2}{r^2} + \frac{c_3}{r} + c_4} .
\end{align}
Performing a Taylor expansion about the point $z \defeq \frac{1}{r} = 0$ to represent the limit $r\to\infty$ leads to the expression
\begin{align}
	f(v,r) \big\vert_{z=0} = \frac{1}{c_4} - \left( \frac{ c_3 + 3c_4 r_{-} + c_4 r_{+}}{c^2_4} \right) \frac{1}{r} + \mathcal{O} \left( \frac{1}{r^2} \right) .
\end{align}
Therefore, to recover the Vaidya form of the metric in the asymptotic limit, we must have 
\begin{align}
	c_4 &= 1 \label{app:eq:ieRBH.coeffc4} , \\
	r_+ &= \frac{c_3 + 3c_4r_{-} + c_{4}r_{+}}{c^2_{4}} \; \; \Rightarrow \; \; c_3 = -3r_{-} \label{app:eq:ieRBH.coeffc3} ,
\end{align}
where Eq.~\eqref{app:eq:ieRBH.coeffc4} was used to obtain the rightmost equality in Eq.~\eqref{app:eq:ieRBH.coeffc3}. Another requirement for nonsingular black holes is that their center be devoid of singularities (as implied by their name). Mathematically, this regularity manifests itself through the non-divergence of spacetime curvature at $r=0$, which can be tracked by evaluating the relevant curvature scalars. Performing a series expansion of the Ricci scalar $R \defeq \sg^{\mu\nu} R_{\mu\nu}$ about this point results in
\begin{align}
	R &= \left( 1 - g(v,0) r^3_{-} r_{+} \right) \frac{2}{r^2} + \Big[ g(v,0) \left( r_- + 3 r_+ \right) \nonumber \\
	& \qquad  \; - r_- r_+ \left(\partial_r g\right) \big\vert_{r=0} \Big] \frac{6 r_-^2}{r} + \mathcal{O} \left(r^0\right) .
\end{align}
To avoid divergences at the center, the coefficients of the first two terms of this expansion must vanish. Consequently, the regularity requirement prescribes the following expressions for the two lowest-order coefficients\footnote{Note that the same result can be obtained by evaluating other scalar curvature invariants, e.g.\ the Kretschmann scalar $K \defeq R_{\mu\nu\rho\sigma} R^{\mu\nu\rho\sigma}$.} of the polynomial $1/g(v,r)$:
\begin{align}
	g(v,0) = \frac{1}{r^3_{-}r_{+}} \; \;  & \Rightarrow \; \;  c_0 = r^3_{-}r_{+} , 
	\label{app:eq:ieRBH.coeffc0} \\
	\left( \partial_r g \right) \big\vert_{r=0} = \frac{r_{-}+3r_{+}}{r^4_{-}r^2_{+}} \; \; & \Rightarrow \; \; c_1 = -r^2_{-}(r_{-} + 3r_{+}) . 
	\label{app:eq:ieRBH.coeffc1}
\end{align}
Additionally, to avoid divergences in $f(v,r)$, the equation $D(v,r)\defeq1/g(v,r)=0$ must not have real solutions [cf.\ Eq.~\eqref{app:eq:ieRBH.fvr}]. This allows us to determine the remaining coefficient $c_2$. Substituting Eqs.~\eqref{app:eq:ieRBH.coeffc4}, \eqref{app:eq:ieRBH.coeffc3}, \eqref{app:eq:ieRBH.coeffc0}, and \eqref{app:eq:ieRBH.coeffc1} into Eq.~\eqref{app:eq:ieRBH.gen.gvr} yields
\begin{align}
	D(v,r) = r_-^3 r_+ - r_-^2 (r_- + 3 r_+) r + c_2 r^2 - 3 r_- r^3 + r^4 . 
\end{align}
This expression can be rewritten as
\begin{align}
	D(v,r) &= r^2 \left(r - \frac{3r_{-}}{2}\right)^2 + \left(c_2 - \frac{15}{4}r^2_{-} - \frac{9}{4} r_{-} r_{+} - \frac{r^3_{-}}{4r_{+}}\right)\nonumber \\
	& \qquad + r^3_{-}r_{+} \left[1 - \frac{1}{2}r \left(\frac{3}{r_{-}} + \frac{1}{r_{+}}\right)\right]^2 .
\end{align}
Since the first and the third term are non-negative and they cannot be zero simultaneously, it suffices to require that the second term be non-negative, i.e.\
\begin{align}
	c_2 - \frac{15}{4} r^2_{-} - \frac{9}{4} r_{-} r_{+} - \frac{r^3_{-}}{4r_{+}} \geqslant 0 ,
	\label{app:eq:ieRBH.coeffc2.ineq}
\end{align}
to ensure that the denominator $D(v,r)$ has no real roots. This non-strict inequality guarantees the positivity of the function $g(v,r)$. We can rewrite Eq.~\eqref{app:eq:ieRBH.coeffc2.ineq} as 
\begin{align}
	c_2 = \tilde{c}_2 + \frac{15}{4} r^2_{-} + \frac{9}{4} r_{-} r_{+} + \frac{r^3_{-}}{4r_{+}} \label{app:eq:ieRBH.coeffc2.c2t}
\end{align}
for some  $\tilde{c}_2 \geqslant 0$, which corresponds to the expression in Eq.~\eqref{eq:ieRBH.c2coeff}. A more explicit form can be obtained from the requirement that the MS mass $C(v,r)/2$ be positive near the black hole center. From the expansion of $C(v,r)$ about $r=0$,
\begin{align}
	C(v,r) = \frac{c_{2} - 3r^2_{-} - 3r_{-}r_{+}}{r^3_{-}r_{+}} r^3 + \mathcal{O}\left(r^4\right) ,
\end{align}
it then follows that $c_2 > 3r^2_{-} + 3r_{-} r_{+}$, which implies that the coefficient $c_2$ can be written as
\begin{align}
	c_2 = 3 r^2_{-} + 3r_{-} r_{+} + \gamma_2 , \quad \gamma_2 > 0 .
\end{align} 
The smallest possible length scale is given by $r_{-}$, which is a plausible choice, although not unique. We use this particular choice in our numerical calculations throughout the paper. In combination with Eq.~\eqref{app:eq:ieRBH.coeffc2.c2t}, it leads to the expression
\begin{align}
	\tilde{c}_2 &= \frac{r_{-}}{4r_{+}} \left( 3 r^2_{+} + r_{-} r_{+} - r^2_{-} \right) .
	\label{app:eq:ieRBH.coeffc2t.expl}
\end{align}
	
\section{Energy-momentum tensor components in the orthonormal frame} \label{app:EMTcomp.ONF}
The EMT components $T_{\mu\nu}$ associated with the general spherically symmetric metric of Eq.~\eqref{eq:gen.metr} are specified through the following relations: 
\begin{align}
	T_{00} &= \frac{e^{h_{+}}}{8 \pi r^2} \left(-e^{h_{+}} f \left( -1 + f + r \partial_{r} f \right) - r \partial_{v} f \right) , 
	\label{app:eq:T00} \\
	T_{01} &= \frac{e^{h_{+}}}{8 \pi r^2} \left( -1 + f + r \partial_{r} f \right) , 
	\label{app:eq:T01} \\
	T_{11} &= \frac{1}{4 \pi r} \partial_{r} h_{+} . 
	\label{app:eq:T11}
\end{align}
To obtain the relevant expressions needed for our calculation in Subsec.~\ref{subsec:EMTclass}, we first write the orthonormal tetrad vectors 
\begin{align}
	e^{\mu}_{\hat{0}} &= \left( e^{-h_{+}} f^{-1/2}, 0, 0, 0 \right) , \\
	e^{\mu}_{\hat{1}} &= \left( e^{-h_{+}} f^{-1/2}, f^{1/2}, 0, 0 \right) , \\
	e^{\mu}_{\hat{2}}& = \left( 0, 0, \frac{1}{r}, 0 \right) , \quad e^{\mu}_{\hat{3}} = \left( 0, 0, \frac{1}{r\sin{\theta}}, 0 \right) .
\end{align}
To obtain explicit expressions for the orthonormal EMT components from those given in Eqs.~\eqref{app:eq:T00}--\eqref{app:eq:T11}, we use the transformation
\begin{align}
	T_{\hat{a}\hat{b}} = e^{\mu}_{\hat{a}} e^{\nu}_{\hat{b}} T_{\mu\nu} .
\end{align}
The resulting orthonormal EMT components are
\begin{align}
	T_{\hat{0}\hat{0}} &= - \frac{f \left( - 1 + f + r \partial_{r} f \right) + e^{-h_{+}} r \partial_{v} f}{8 \pi r^2 f} , \\
	T_{\hat{0}\hat{1}} &= - \frac{e^{-h_{+}} \partial_{v} f}{8 \pi r f} , \\
	T_{\hat{1}\hat{1}} &= - \frac{-f \left[ -1 + r \partial_{r} f + f \big(1 + 2r (\partial_{r} h_{+}) \big) \right] + e^{-h_{+}} r \partial_{v} f}{8 \pi r^2 f} , \\
	T_{\hat{2}\hat{2}} &= T_{\hat{3}\hat{3}} = \frac{1}{16 \pi r} \Bigg[
	\partial_{r} f \big(2 + 3 r \partial_{r} h_{+} \big) + 2 f \Big[ \partial_{r} h_{+} \nonumber \\
	& \quad + r \big(\partial_{r} h_{+} \big)^2 + r \big(\partial^2_{r}h_{+}\big) \Big] + r \big(\partial^2_{r} f + 2 e^{-h_{+}} \partial_{r} \partial_{v} h_{+} \big)
		\Bigg] .
\end{align}

\end{document}